\documentclass[10pt]{article}
\usepackage{amssymb}
\usepackage{amsmath}
\usepackage{graphicx}
\usepackage{a4wide}

\newcommand{\Nloc}{N_\text{loc}}
\newcommand{\Nseg}{N_\text{seg}}
\newcommand{\bNloc}{\bar{N}_\text{loc}}
\newcommand{\bNseg}{\bar{N}_\text{seg}}
\newcommand{\tNloc}{N^*}

\begin{document}

\title{Supplemental Material for Comment on ``Excitons in Molecular Aggregates with L\'evy Disorder: Anomalous Localization and Exchange Broadening of Optical Spectra'', Phys.\,Rev.\,Lett.~109, 259701}
\author{Agnieszka Werpachowska}
\date{October 23, 2011}


\maketitle

\begin{abstract}
In their Letter ``Excitons in Molecular Aggregates with L\'evy Disorder: Anomalous Localization and Exchange Broadening of Optical Spectra'', Eisfeld~\textit{et al.}\ predicted the existence of exchange broadening and blue-shift of the absorption band and a non-universal disorder scaling of the localisation length of absorption band excitons in J-aggregates with heavy-tailed disorder, which they contrasted with the previously analysed Gaussian and Lorentzian case. The observations were explained by another localisation mechanism, the chain segmentation by outliers in heavy tails of the disorder distribution. We argue that the previously known theory does not break down and consequently anticipates the properties of the absorption band investigated in the commented work.
\end{abstract}

\section{Introduction}

In their Letter~\cite{prl10}, Eisfeld~\textit{et al.}\ predicted the existence of exchange broadening and blue-shift of the optical absorption band and a non-universal disorder scaling of the localisation length $\Nloc$ of the excitons therein, in J-aggregates with uncorrelated site disorder governed by the heavy-tailed L\'evy distribution (with stability index $\alpha=\frac{1}{2}$). They contrasted it with the previously analysed Gaussian ($\alpha=2$) and, also heavy-tailed, Lorentzian ($\alpha=1$) case~\cite{prb95,prb09}. The observations were explained by another localisation mechanism, namely the chain segmentation due to high concentration of outliers in heavy tails of the disorder distribution, and its interplay with the usual localisation in effective potential wells created by typical random site energies.

Following the Letter, we analyse the disordered exciton chains of $N=200$ sites described by the Hamiltonian $H_{nm} = \delta_{nm} E_n - (\delta_{n,n+1} + \delta_{n,n-1}) J$, where $E_n$ are random site energies drawn from the L\'evy distribution\footnote{We have tested our results for different numbers of the disorder realisations---from one hundred thousand to five million.} and $J=1$ is the nearest-neighbour interaction strength. In Sec.~\ref{sec:universality}, we show that L\'evy disorder does not break the universality of localisation length distribution in the absorption band observed for the Gaussian and Lorentzian case~\cite{prb09}, pointing out the error in the Authors' calculations in Fig.\,3 of Ref.\,\cite{prl10}. The recalculated half width at half maximum (HWHM) scaling and blue-shift result from the previously known theory~\cite{prb09}, similarly to the exchange broadening (whose range of occurrence was incorrectly assessed in Fig.\,2 of Ref.\,\cite{prl10}), as we argue in Secs.~\ref{sec:hwhm}--\ref{sec:blue-shift}. We conclude that theory~\cite{prb09} provides a full statistical picture of the absorption band properties in systems with different disorder types, taking into account both microscopic mechanisms of segmentation and localisation in potential wells.

\section{Universality of the localisation length scaling}
\label{sec:universality}

The Letter refers to another article~\cite{prb09}, which investigates the localisation properties of similar systems with Gaussian and Lorentzian disorder. In particular, it is shown there that the average localisation length $\bNloc$ in the absorption band scales with the disorder in the same way as the typical localisation length
\begin{equation}
\label{eq:lawprb09}
\tNloc \sim \sigma^{-\alpha/1 + \alpha} \ .
\end{equation}
The standard deviation $\delta \Nloc$ follows a similar scaling with almost the same exponent, implying that the shape of the distribution is universal, $\delta \Nloc/\bNloc \approx \text{const.}$ (see Figs.~3 and~4 of Ref.\,\cite{prb09}). In the Letter, the Authors predict the breakdown of this universality for the case of L\'evy disorder, as shown in Fig.\,3 of Ref.\,\cite{prl10}.

\begin{figure}[htbp]
\centering
\includegraphics[scale=0.9]{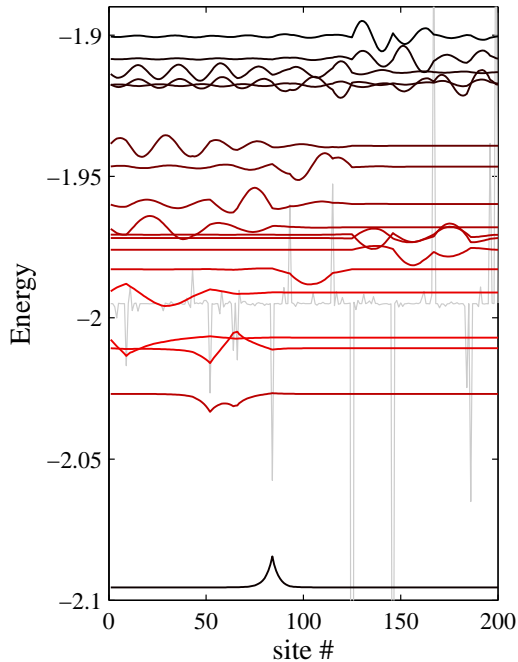}
\caption{Typical set of states from the energy range $[-2.1,-1.9]$ for L\'evy disorder and $\sigma=0.001$ with energy landscape in the background.}%
\label{fig:typset}%
\end{figure}

We argue that the reported non-universality of the localisation length distribution does not arise from the properties of L\'evy disorder, but is a consequence of the wrong choice of energies included in the calculations. The constant range $\epsilon \in [-2.1,-1.9]$ does not adjust to disorder-induced scaling and shifts of the absorption band, like it was assured by using the $\tilde{\epsilon} \in [-0.1,0]$ range (i.e.~$\epsilon$ range scaling with $\sigma$) for other disorder types in Ref.\,\cite{prb09}. This causes, for low $\sigma$ values, a too wide variety of states to be included in the calculation, as shown in Fig.\,\ref{fig:typset}. As a consequence, the standard deviation $\delta \Nloc$ disproportionately increases and the ratio $\delta \Nloc/\bNloc$ is no longer constant. The same effect is observed for Gaussian and Lorentzian disorder. The redundant states are also responsible for the deformation of the localisation length distribution shown in the inset of Fig.\,3 of Ref.\,\cite{prl10}. 

The correct energy range accounts for the scaling and shift of the absorption band with disorder~\cite{prl08} according to the following formula:
\[
\tilde{\epsilon} = \frac{\epsilon - \epsilon_b + a \sigma^{\alpha'}}{b \sigma^{\alpha'}} \ ,
\]
where $\tilde{\epsilon}$ is a universal energy variable independent of $\sigma$ and $\epsilon_b$ is the band-edge energy, equal $-1.99976$ in the analysed system (deviating from the value of $-2$ due to the energy quantisation in a finite chain). The values of constants $a$ and $b$ and exponent $\alpha'$ for Gaussian and Lorentzian diagonal disorder with nearest-neighbour interactions and other systems were derived in Refs.\,\cite{prb09,klugkist} from the scaling analysis of the Lifshitz tail of the joint distribution $\mathcal G(\epsilon,\mu)$ of energy $\epsilon$ and transition dipole moment $\mu$. Performing this analysis for L\'evy disorder, we have derived the values $a = -0.23$, $b = 5$ and $\alpha'=2/3$. (As long as we scale the energy range with $\sigma^{\alpha'}$, the results are not very sensitive to the choice of $a$ and $b$ parameters.)

\begin{figure}[htbp]%
\centering
\includegraphics[scale=0.7]{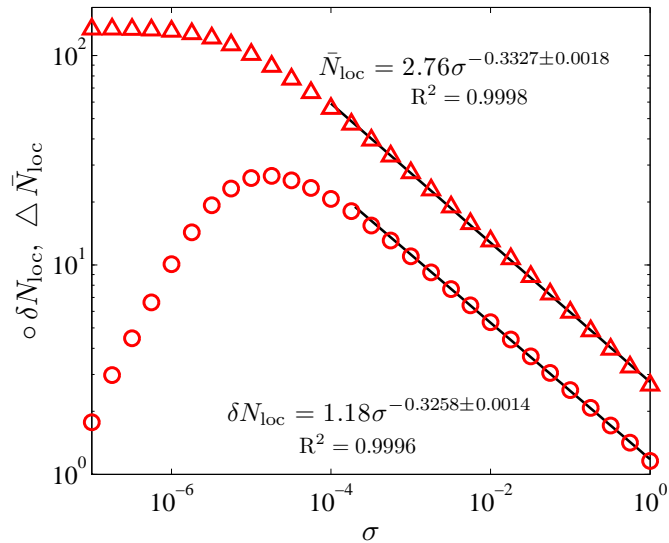}
\caption{Universal scaling of the average and standard deviation of the localization length distribution in a chain with L\'evy disorder.}%
\label{fig:scalinglevylog}%
\end{figure}

Using the correct energy ranges we obtain universal scaling of the $\Nloc$ distribution for a wide range of L\'evy disorder strengths, breaking down only in the limit $\sigma \to 0$ due to the convergence of the localisation length distribution towards a single value $\Nloc = 134$ (for chain size $N=200$) characterising unperturbed states (Fig.\,\ref{fig:scalinglevylog}). The average localisation length $\bNloc$ is proportional to $\sigma^{-1/3}$, in accordance with the typical localization length scaling law~\eqref{eq:lawprb09}. The standard deviation $\delta\Nloc$ scales with almost the same exponent as the first moment, $\sigma^{-0.324}$. The difference of exponents, although not zero, is of the same order as in the case of Gaussian and Lorentzian disorder reported in Ref.\,\cite{prb09}, and much smaller than in the commented Letter. This underlines the importance of using the correct energy range scaling in the characterisation of the localisation length distribution in the absorption band.

\begin{figure}[htbp]%
\centering
\includegraphics[scale=0.7]{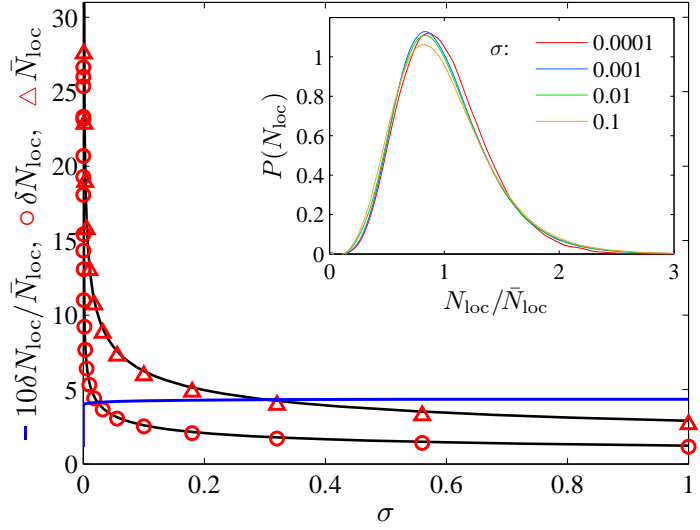}
\caption{Disorder scaling of the average and standard deviation of the localization length distribution in a chain with L\'evy disorder, and their constant ratio. Inset: renormalised localization length distributions for few different values of disorder strength.}%
\label{fig:avglen200}%
\end{figure}

As demonstrated in Fig.\,\ref{fig:avglen200}, the ratio $\delta \Nloc/\bNloc$ remains stable. The inset demonstrates that the renormalised localisation length distributions do not depend on $\sigma$ and thus have universal shapes. Consequently, the typical $\tNloc$ and average $\bNloc$ localisation lengths follow the law~\eqref{eq:lawprb09} with the same scaling exponent.

\section{Segmentation scaling law}
\label{sec:seg-scale}

\begin{figure}[htbp]%
\centering
\includegraphics[scale=0.62]{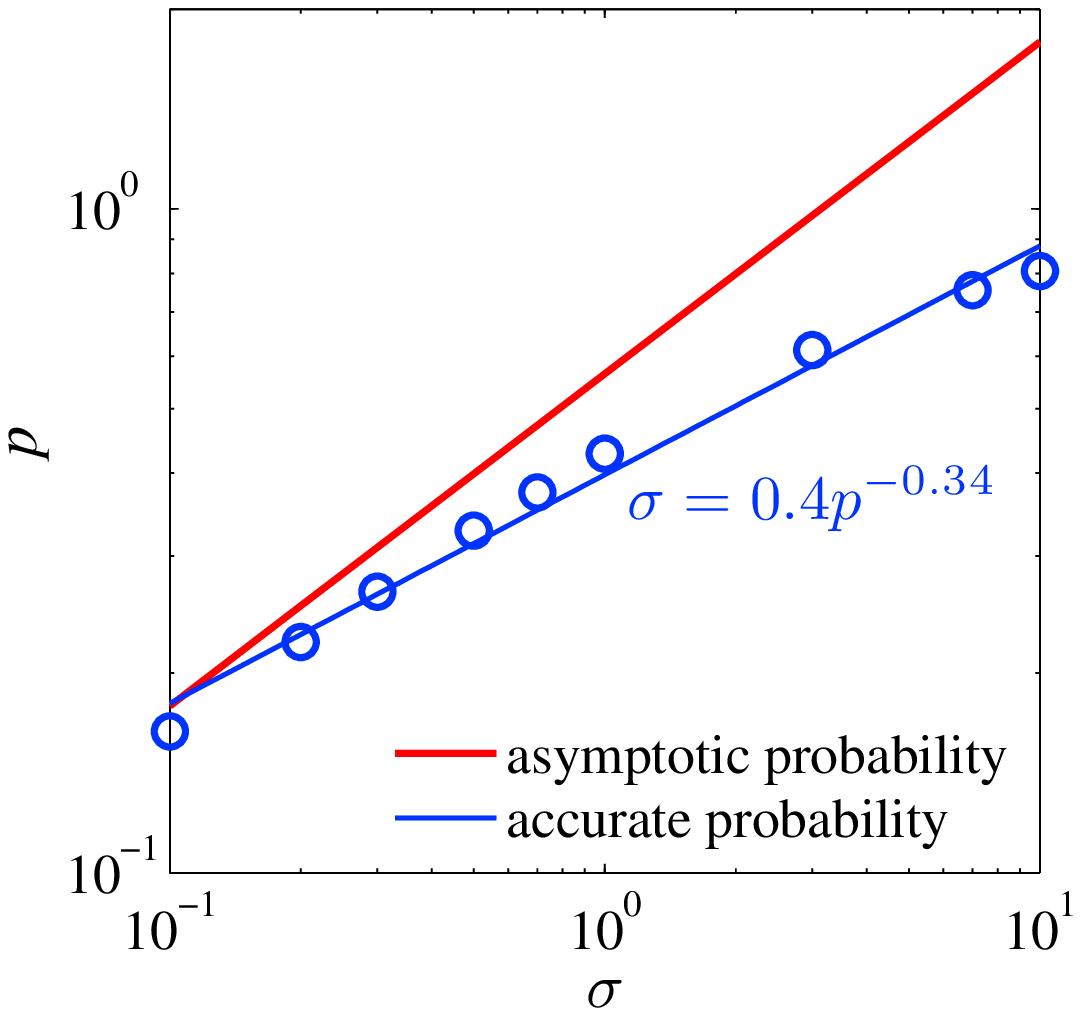}\ \ \includegraphics[scale=0.62]{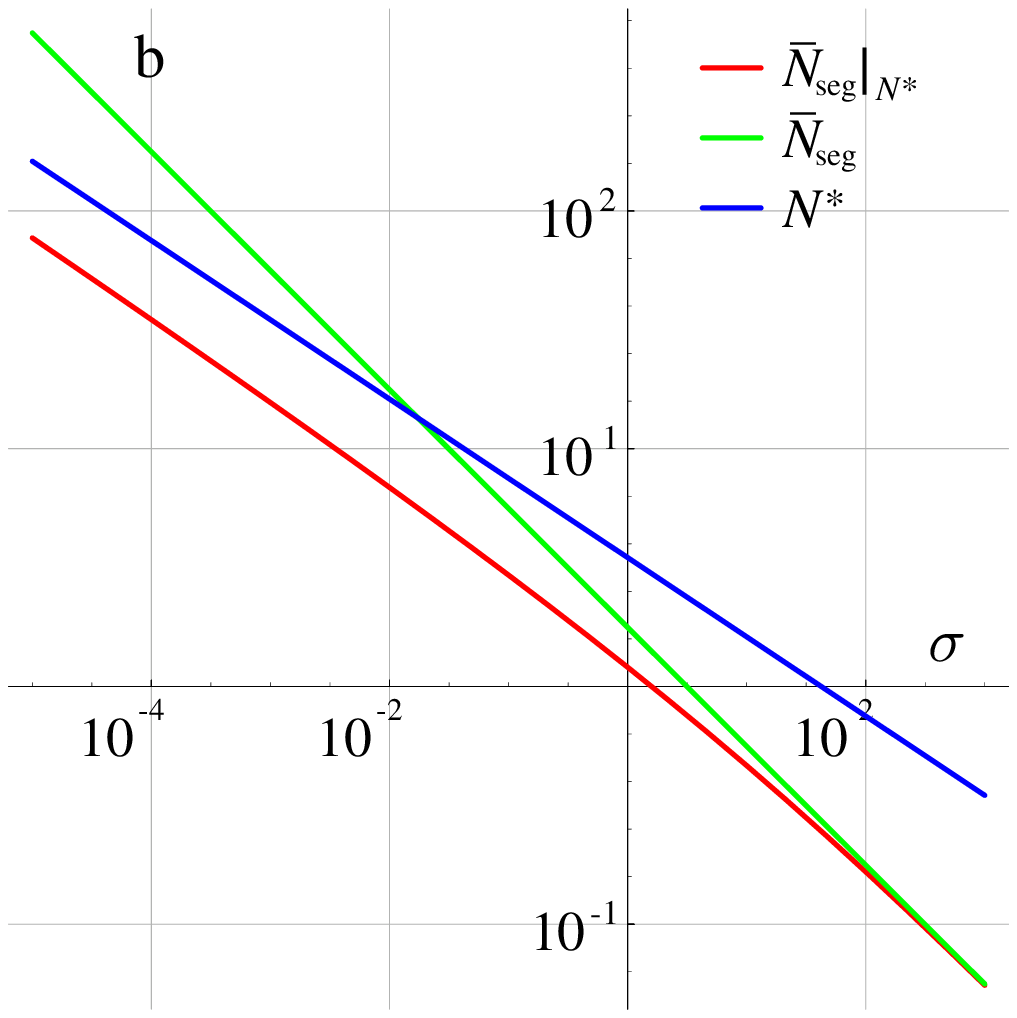}
\caption{a) Breakdown of the asymptotic scaling relation between $p$ and $\sigma$ for $\alpha=\frac{1}{2}$. b) Segmentation scaling law with and without the constraint $\Nseg < \tNloc$ compared to the $\tNloc$ scaling. }%
\label{fig:psigma}%
\end{figure}

The Letter explains the predicted breakdown of the localisation length scaling by the chain segmentation mechanism induced by outliers, i.e.~site energies $|E_n| > 2$, in heavy tails of L\'evy disorder distribution. Although we have shown in Sec.~\ref{sec:universality} that the universality holds also in this case, it is worth to discuss the effect of this mechanism on the properties of absorption band. It introduces another characteristic length, the average length of chain segments cut out by neighbouring outliers, following the scaling law (Eq.\,4 of Ref.\,\cite{prl10}),
\begin{equation}
\bNseg \sim \sigma^{-\alpha} \ ,
\label{eq:nseg}
\end{equation}
It arises from the fact that for the disorder uncorrelated between sites, $\bNseg = 1/p$ (assuming that the chain size is much larger than $1/p$), where $p$ is the probability of the occurrence of an outlier on a single site. Using the asymptotic form of disorder probability density function, for $\alpha < 2$,
\begin{equation}
f(E_n) \approx \frac{\sigma^\alpha \sin( \pi \alpha / 2 ) \Gamma(\alpha+1)}{\pi |E_n|^{1+\alpha}}
\label{eq:asympt}
\end{equation}
to derive the scaling $p = P(|E_n| > 2) \sim \sigma^\alpha$, one obtains~\eqref{eq:nseg}. For L\'evy disorder, this asymptotic scaling relation between $p$ and $\sigma$ breaks down already for $\sigma \gtrsim 0.1$ and $p$ scales with $\sigma$ according to exponents higher than~$-\frac{1}{2}$ (Fig.\,\ref{fig:psigma}a). Thus, the approximate expression~\eqref{eq:asympt} for the asymptotic behaviour of L\'evy distribution cannot be used to calculate the outlier occurrence probability for larger $\sigma$ values, and consequently the segmentation scaling law~\eqref{eq:nseg} cannot be valid in this regime.

The scaling of $\bNseg$ cannot be applied to the scaling of $\bNloc$, since different segments contribute different numbers of states to the absorption band: short segments (belonging to group (iii) in Ref.\,\cite{prl10}) contribute only a single state, while the longer ones contribute two or more. Hence, $\bNloc$ is in general larger than $\bNseg$ and scales with a different exponent.

If we consider just the short segments from group (iii) (with lengths $\Nseg < \tNloc$), which contain one absorption band state each, the average of their lengths indeed equals the average of localisation lengths of the absorption band states they contain. However, due to the above constraint its scaling will be influenced by the scaling of $\tNloc$ in the following way:
\[
\bNseg|_{\tNloc} = \mathbb{E}[\Nseg | \Nseg < \tNloc] = \frac{\int_0^{\tNloc} N e^{-N/\bNseg} dN}{\int_0^{\tNloc} e^{-N/\bNseg} dN} = \frac{\bNseg - e^{-\frac{\tNloc}{\bNseg}} ( \bNseg + \tNloc )}{1 - e^{-\frac{\tNloc}{\bNseg}}}\ .
\]
Since $\alpha > \frac{\alpha}{1 + \alpha}$, for low $\sigma$ we have $\tNloc \ll \bNseg$ and $\bNseg|_{\tNloc}$ scales approximately like $\tNloc$. Using the prefactors from Eqns.~3 and 4 in Ref.\,\cite{prl10}, we obtain that, with good accuracy, for L\'evy disorder $\bNseg|_{\tNloc} \sim \sigma^{-0.36}$ for $\sigma \in [10^{-5}, 1]$, as demonstrated in Fig.\,\ref{fig:psigma}b. It contrasts with the segmentation scaling law~\eqref{eq:nseg}, which predicts scaling with exponent $-\frac{1}{2}$.
On the other hand, for large $\sigma$ we have $\tNloc \gg \bNseg$ and the scaling $\bNseg|_{\tNloc}$ will approach that of $\bNseg$. However, in this regime $\bNseg$ no longer scales with exponent $-\alpha$, as we have shown above.

The law~\eqref{eq:nseg} could become relevant for very strong disorder, when almost all states belong to the absorption band. Then, one can assume that each segment of length $\Nseg$ contributes $\Nseg$ states with lengths $\Nseg$, obtaining
\begin{equation}
\bNloc \sim \frac{\int_0^\infty N^2 \,e^{-N/\bNseg} \mathrm{d}N}{\int_0^\infty {N}\, e^{-N/\bNseg} \mathrm{d}N} = 2 \bNseg \sim \sigma^{-\alpha}\ .
\label{eq:bNlocSimbNseg}
\end{equation}
However, the outlier occurrence probability scaling law $p \sim \sigma^{-\alpha}$ breaks down in this case (Fig.\,\ref{fig:psigma}a). Moreover, for such strong localisation the power law scaling saturates due to the natural lower bound $\Nloc \ge 1$.

\section{HWHM and exchange broadening}
\label{sec:hwhm}

The Letter derives a quantitative explanation for the exchange broadening of absorption band in the presence of L\'evy disorder from the interplay between the two mechanisms for localising the exciton states: the localisation on effective potential wells created by random site energies and the chain segmentation by outliers. The Authors associate them with scaling laws for the HWHM with $\sigma^{2\alpha/1+\alpha}$ and $\sigma^{\alpha}$ (which gives $\sigma^{2/3}$ and $\sigma^{1/2}$ for $\alpha=\frac{1}{2}$), respectively, obtained by inserting $N^\star$ or $\bNseg$ in place of $\Nloc$ in the expression
\begin{equation}
\sigma^* = \sigma \Nloc^{\frac{1-\alpha}{\alpha}}
\label{eq:avgdis-simple}
\end{equation}
for the site-averaged disorder. This follows from their assumption that each of the two localisation mechanisms has its own length scale: $\tNloc$ for localisation in potential wells, and $\bNseg$ for segmentation. From the numerically obtained scaling exponent $0.6\pm0.03$ they draw the conclusion that both mechanisms contribute equally to this effect. 

\begin{figure}[htbp]%
\centering
\includegraphics[scale=0.635]{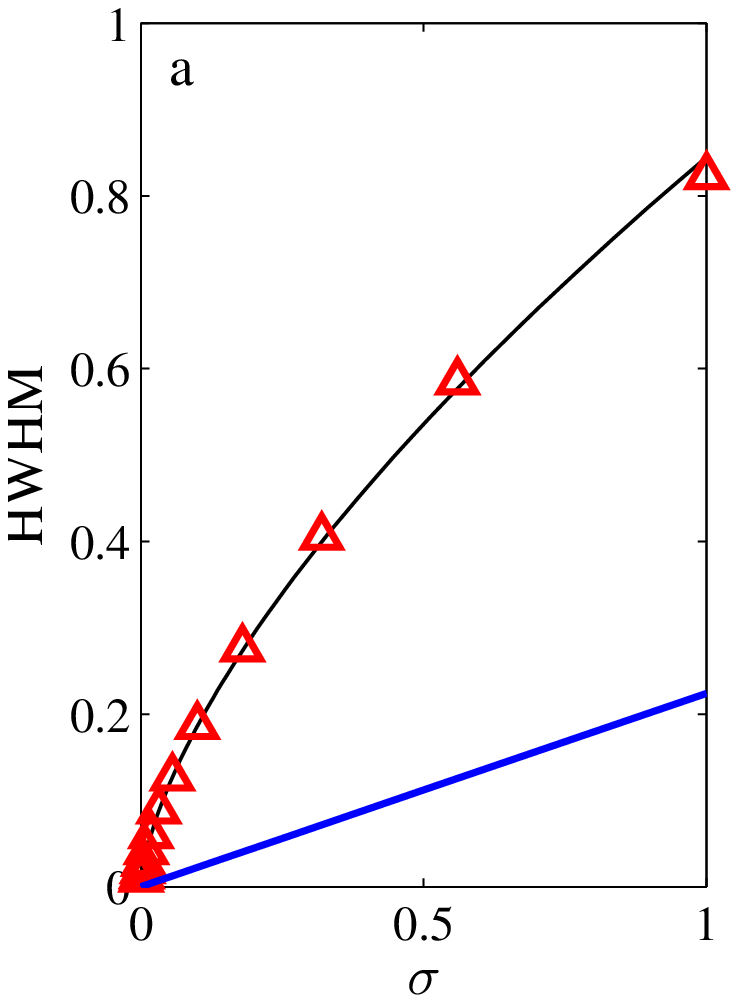}\includegraphics[scale=0.635]{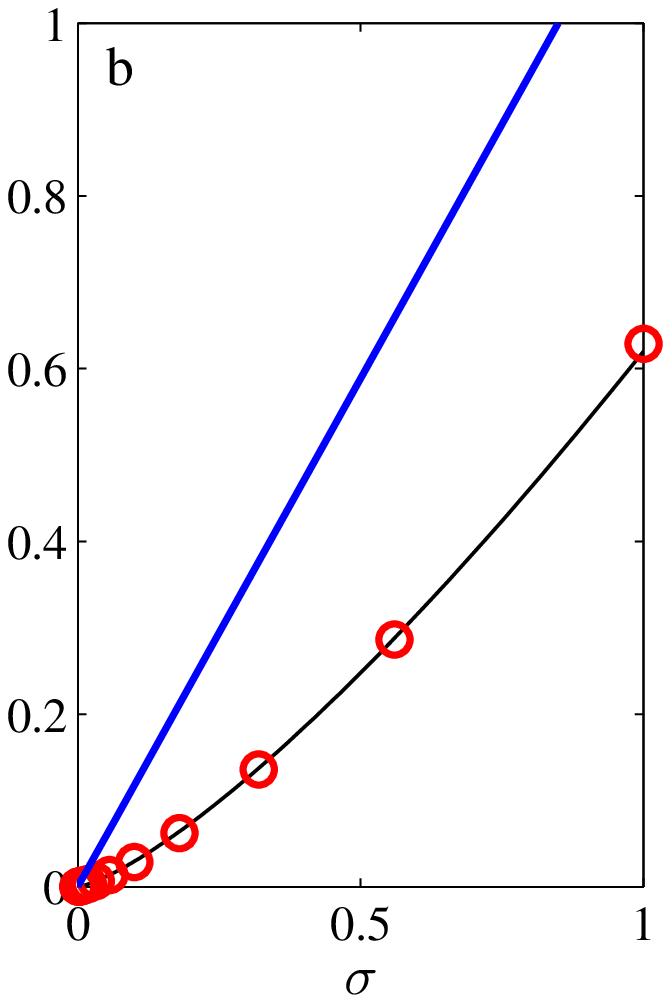}\includegraphics[scale=0.635]{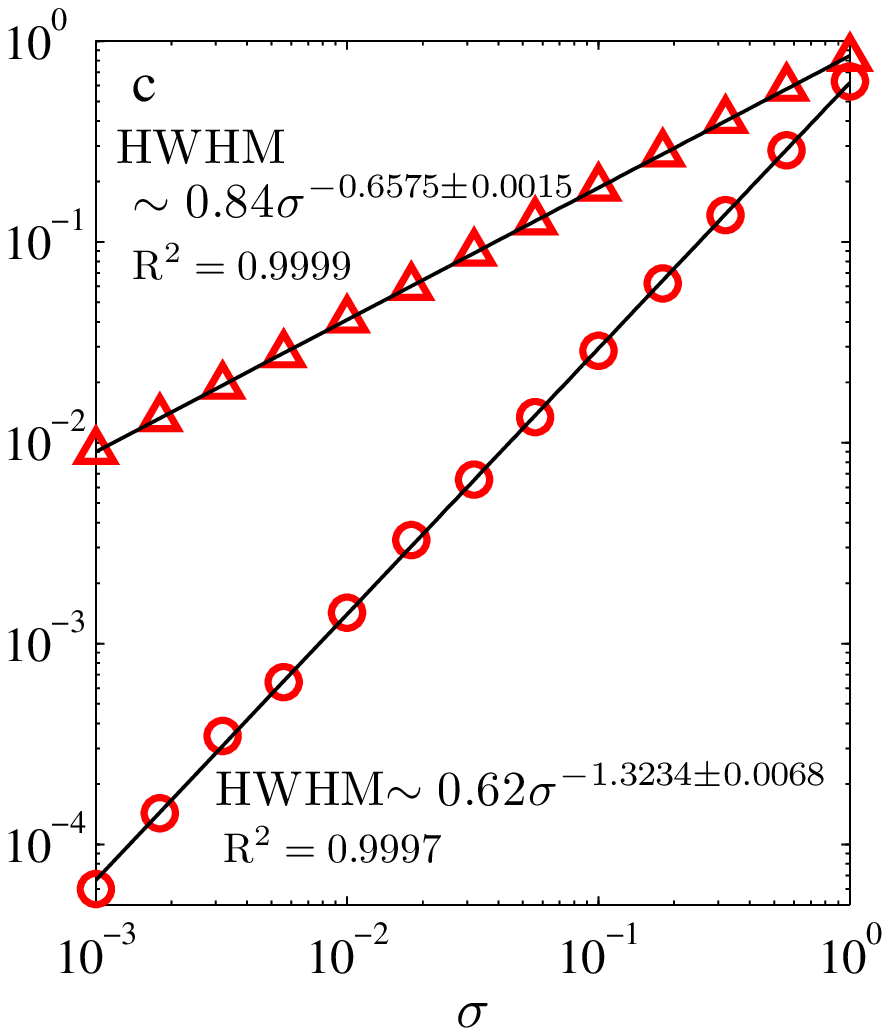}
\caption{Comparison of the absorption band HWHM (red markers) fit by the power law~\eqref{eq:lawprb09} (black line) with the bare disorder distribution HWHM (blue line) for L\'evy (a) and Gaussian (b) distributions. c) The power law fits for L\'evy (triangles) and Gaussian (circles) disorder; logarithmic scale was chosen to give more accurate results when fitting a power function. The presented fitting error does not take into account the error of the estimation of HWHM values from the absorption spectrum plots for each $\sigma$.}%
\label{fig:hwhm}%
\end{figure}

We have recalculated the HWHM of the absorption band for L\'evy disorder, taking particular care to obtain good accuracy for low $\sigma$ values, for which the absorption band becomes very narrow. To account for this, we have used a smaller histogram bin size than for higher $\sigma$ values. As presented in Fig.\,\ref{fig:hwhm}c, the obtained scaling exponent $0.657\pm 0.002$ is in good agreement with the $\sigma^{2/3}$ power law for HWHM, which follows from the law~\eqref{eq:lawprb09} for typical localisation length. This is due to the universality of the distribution of $\Nloc$, demonstrated in Sec.~\ref{sec:universality}, and the fact that $\tNloc$ is a typical localisation length of all absorption band states, determined by both localisation mechanisms.

It is also important to note that the discussion of Fig.\,2 in Ref.\,\cite{prl10} contains an error: to check for exchange narrowing or broadening, the HWHM of the absorption band is compared with the $\sigma$ value. Instead, it should be compared with the bare disorder HWHM, equal $\sqrt{2 \ln 2} \sigma$ for Gaussian and $0.224 \sigma$ (obtained numerically) for L\'evy disorder (Fig.\,\ref{fig:hwhm}\,a,b). When the correct function is used, the HWHM curve for $\alpha=\frac{1}{2}$ exhibits exchange broadening for all $\sigma$ values considered---not over the $[0,0.6]$ range only.

\section{Blue-shift of the absorption band}
\label{sec:blue-shift}

The Authors used the segmentation mechanism to explain the blue-shift of the absorption band for L\'evy disorder as opposed to the Gaussian one, for which a red-shift is observed, and the Lorentzian, where no shift occurs. However, the outliers occur already for the last type of disorder~\cite{prb09}, which also has a heavy-tailed distribution with infinite variance.\footnote{It may be relevant to note that all heavy-tailed $\alpha$-stable distributions, i.e.~those with $\alpha < 2$, have divergent second moment, while those with $\alpha < 1$ have also undefined mean (even if they are symmetric around some value). To regularise the expectation values involving the latter, we can replace the means with the principal values calculated numerically using antithetic sampling.} This suggests that the blue-shifts of the absorption band position are not controlled by the segmentation mechanism only. We argue that the full explanation of the shifts of the absorption band for different types of disorder needs to be formulated on the grounds of the previously known theory~\cite{prb09}.

\begin{figure}[htbp]
\centering
\includegraphics[scale=0.8,trim=1 1 1 1,clip=true]{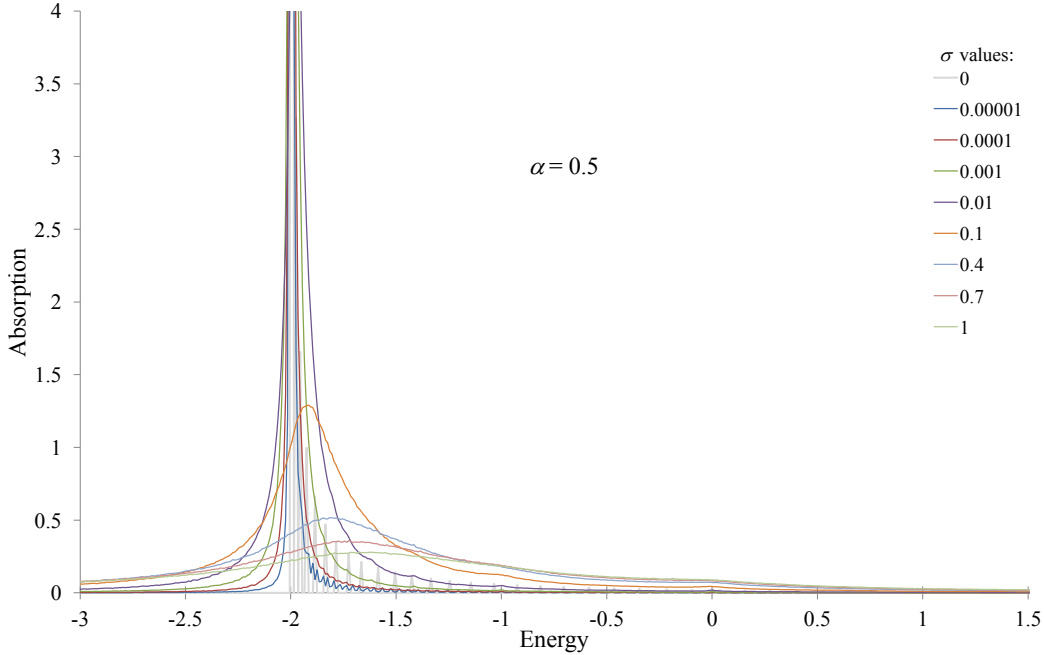}
\caption{Blue-shift and broadening of the absorption spectrum due to L\'evy disorder. The quantisation of the absorption spectrum in an unperturbed system is visible, smearing out gradually with growing disorder.}%
\label{fig:bs0.5}%
\end{figure}

The energy of a delocalised state $\varphi$ is perturbed by the disorder by an amount equal to its average value $\sum_n |\varphi_n|^2 E_n$, where $E_n$ is the value of the disorder on site $n$ (bare disorder). The perturbation is distributed with the same index of stability $\alpha$ as the bare disorder, but with a different scale parameter
\begin{equation}
\sigma^* = \sigma \left( \sum_n |\varphi_n|^{2\alpha} \right)^{1/\alpha} .
\label{eq:avgdis}
\end{equation}
For Gaussian disorder, it gives $\sigma^* = \sigma \sqrt{\mathcal{L}}$ (where $\mathcal{L} = \sum_n |\varphi_n|^4$ is the inverse participation ratio of state $\varphi$), for Lorentzian $\sigma^* = \sigma$, while for L\'evy $\sigma^* = \sigma \left\lVert \varphi \right\rVert^2_1$ (where $\left\lVert \varphi \right\rVert_1 = \sum_n |\varphi_n|$ is the $\ell^1$-norm of $\varphi$). Assuming that the state $\varphi$ is distributed uniformly over $\Nloc$ sites, we obtain the previously known, simpler formula~\eqref{eq:avgdis-simple}. Thus, states with different localisation lengths experience different average disorder strengths, which leads to the width and position of the absorption spectrum changing with $\sigma$. The states with larger localisation lengths are mostly optically inactive. In the case of Gaussian disorder, they feel weaker average disorder $\sigma^* \sim \sigma / \sqrt{\Nloc}$ (exchange narrowing) and thus are spread less away from the band centre $\epsilon = 0$ than the optically active states, pushing the absorption band away from it (red-shift). For L\'evy disorder, they feel larger averaged disorder $\sigma^* \sim \sigma \Nloc$ (exchange broadening) and are spread more away from the band centre than the optically active states, pushing the absorption band towards it (blue-shift, Fig.\,\ref{fig:bs0.5}). For Lorentzian disorder, all states are spread equally ($\sigma^* = \sigma$) and no shift (or narrowing/broadening) of the absorption band occurs. 
Apart from that, the states become increasingly localised as $\sigma$ grows, according to the scaling law~\eqref{eq:lawprb09}, which intensifies the above effects.
Thus, in the case of L\'evy disorder the blue-shift of the absorption band follows from the previously known theory~\cite{prb09}.

The presented statistical picture is realised by the two microscopic mechanisms, the localisation of states in potential wells and the chain segmentation. It is their interplay, captured by the formulas~\eqref{eq:lawprb09} and~\eqref{eq:avgdis}, which decides on the shape and position of the absorption band. From the results of the macroscopic theory, we can decipher the microscopic scenarios for different disorder types. For $\alpha > 1$, the optically active states mostly localise in potential wells lowering their potential energy (arising from the disorder) with the growth of $\sigma$, while the effect of the few outliers, squeezing the states on shorter segments and thus pushing up their kinetic energy (arising from the interaction), is negligible. As a result the red-shift is observed. For $\alpha = 1$ the two effects compensate perfectly. For any $\alpha < 1$ the segmentation effect prevails and the energy of states rises, resulting in the blue-shift of the absorption band. The elements of this microscopic picture should not be mixed with the statistical one, described in the previous paragraph.

\section{Conclusions}

We conclude that, although one might expect that the presence of outliers in the heavy tails of L\'evy distribution would affect markedly the properties of the absorption band, the previously known theory, derived and applied to Gaussian and Lorentzian disorder in Ref.\,\cite{prb09} remains valid. The localisation length scaling~\eqref{eq:lawprb09} is a macroscopic law arising from the fact that the typical energetic cost of localising the absorption band states must equal the typical site-averaged disorder experienced by them. This averaged disorder carries the information about the type of its distribution in form of the relation~\eqref{eq:avgdis-simple}. In particular, for L\'evy disorder it predicts $\sigma^* > \sigma$, due to the presence of outliers. On the other hand, the chain segmentation and localisation of states in potential wells are just microscopic mechanisms realising this theory. Contrary to the discussion in the Letter, both mechanisms are associated with a common length scale $\tNloc$ described by theory~\cite{prb09}. 

\begin{figure}[htbp]
\centering
\includegraphics[scale=0.8,trim=1 1 1 1,clip=true]{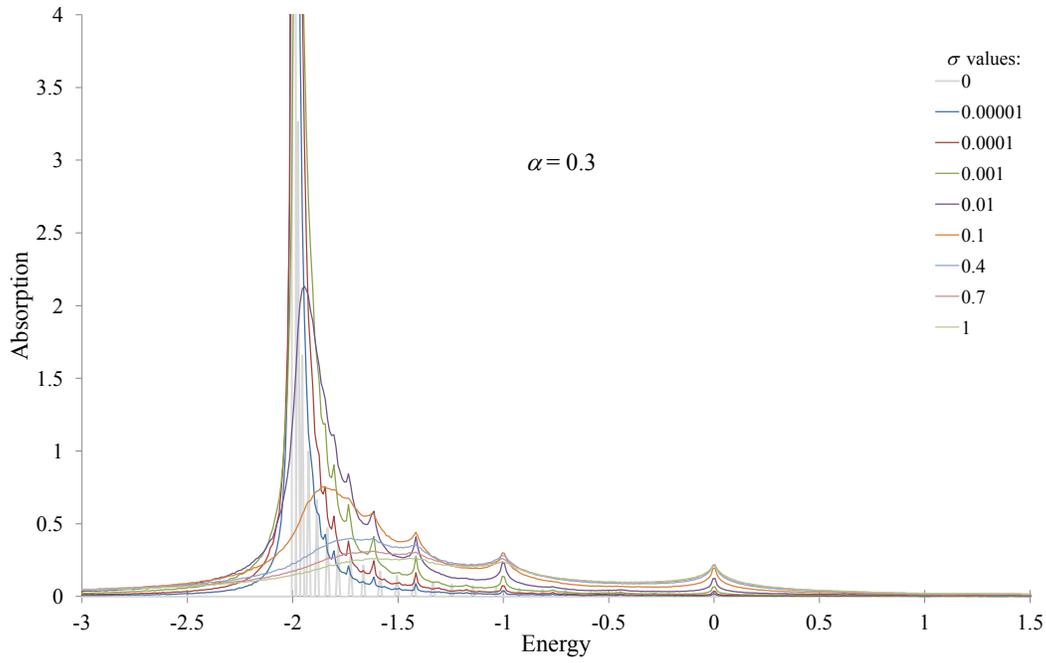}
\caption{Absorption spectrum for $\alpha=0.3$.}%
\label{fig:bs0.3}%
\end{figure}

\begin{figure}[htbp]
\centering
\includegraphics[scale=0.8,trim=1 1 1 1,clip=true]{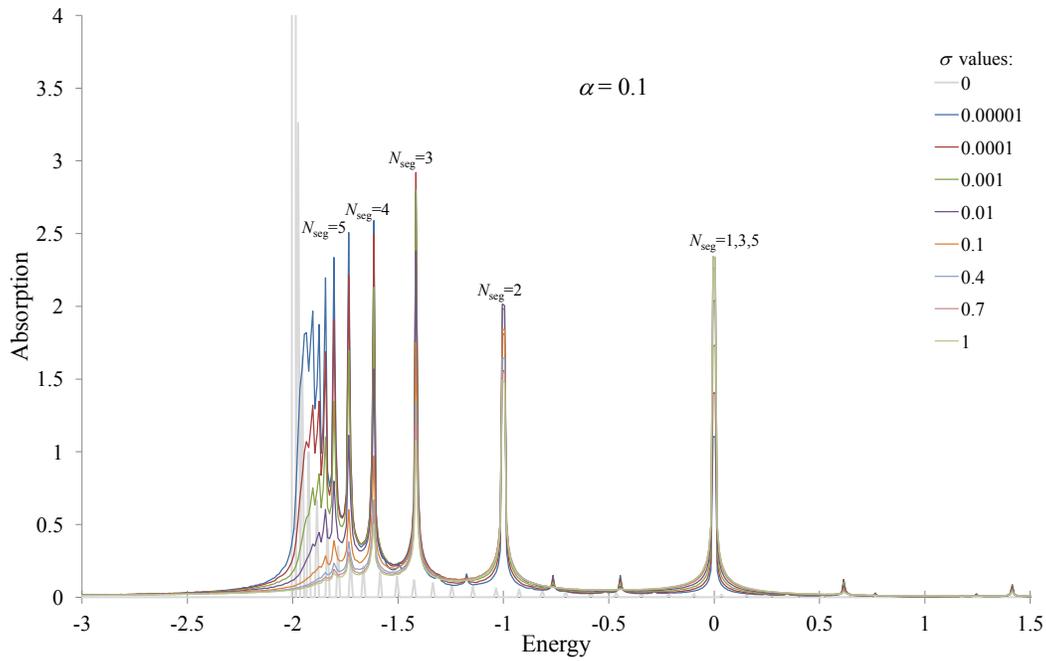}
\caption{Strongly quantised absorption spectrum for $\alpha=0.1$ with marked lengths of short segments $N_\text{seg}$ in the absorption peaks.}%
\label{fig:bs0.1}%
\end{figure}

The unexpected features of the absorption spectrum, i.e.~the peaks corresponding to the shortest chain segments, become more prominent for low stability indices $\alpha$, as predicted by the Authors. In this regime the segmentation mechanism becomes dominant, as the density of outliers increases and outweighs the density of typical site energies. This is due to the fact that as $\alpha \to 0$, the disorder distribution is composed of long heavy tails and a part increasingly centred at zero. Then, the states become strongly localised by the outliers, at the same time feeling much weaker disorder \emph{between} them. As a result, the density of states becomes quantised at fixed energy values corresponding to very short ($\Nloc \lesssim 5$) localisation lengths, and the absorption spectrum splits in a series of peaks. In this regime the HWHM becomes undefined, while the absorption band shift occurs purely through the transfer of the absorption spectrum density between its multiple peaks towards higher energies corresponding to shorter segments, as demonstrated in Figs.~\ref{fig:bs0.3} and~\ref{fig:bs0.1}. These peaks are slightly broadened even in the limit of $\alpha \to0$, because the energy of the states they contain is smeared by the interaction of intrasegment sites with boundary sites containing random values of outliers. For $\sigma \gg 1$, the density of states and the absorption spectrum become concentrated in the $\Nloc = 1$ peak at $\epsilon=0$ only, which will be smeared by the disorder, yielding a monomer spectrum with the shape given by the bare disorder distribution~\cite{prl10}. In this limit, all states participate in absorption, but the relation $\bNseg \sim \sigma^{-\alpha}$ is no longer valid, as we have discussed in Sec.~\ref{sec:seg-scale}.

\begin{thebibliography}{1}

\bibitem{prl10}
A.~Eisfeld, S.~M. Vlaming, V.~A. Malyshev, and J.~Knoester.
\newblock Excitons in molecular aggregates with {L}\'evy disorder: Anomalous
  localization and exchange broadening of optical spectra.
\newblock {\em Phys. Rev. Lett.}, 105(13):137402, 2010.

\bibitem{prb95}
Victor Malyshev and Pablo Moreno.
\newblock Hidden structure of the low-energy spectrum of a one-dimensional
  localized {Frenkel} exciton.
\newblock {\em Phys. Rev. B}, 51:14587, May 1995.

\bibitem{prb09}
S.~M. Vlaming, V.~A. Malyshev, and J.~Knoester.
\newblock Localization properties of one-dimensional {F}renkel excitons:
  {G}aussian versus {L}orentzian diagonal disorder.
\newblock {\em Phys. Rev. B}, 79(20):1--8, 2009.

\bibitem{prl08}
J.~A. Klugkist, V.~A. Malyshev, and J.~Knoester.
\newblock Scaling and universality in the optics of disordered exciton chains.
\newblock {\em Phys. Rev. Lett.}, 100:216403, May 2008.

\bibitem{klugkist}
J.~A. Klugkist.
\newblock {\em Mechanisms for photonic switching in systems of strongly
  interacting dipoles}.
\newblock PhD thesis, Rijksuniversiteit Groningen, 2007.

\end{thebibliography}

\end{document}